\documentclass[12pt]{article}
\usepackage{xcolor,amsmath}
\usepackage{breqn}
\usepackage{cite}
\usepackage{soul}
\usepackage{cancel}

\begin{document}

\title{\bf Exponential correction to Friedmann equations}

\author{{\"{O}zg\"{u}r \"{O}kc\"{u}$^{a,b}$ \thanks{Email: ozgur.okcu@ogr.iu.edu.tr}\hspace{1mm} Ekrem Aydiner$^{a,b}$ \thanks{Email: ekrem.aydiner@istanbul.edu.tr}} \\
$^a${\small {\em Theoretical and Computational Physics Research Laboratory, Istanbul University,
		}}\\{\small {\em Istanbul, 34134, Türkiye}} \\
$^b${\small {\em Department of
	Physics, Faculty of Science, Istanbul University,
		}}\\{\small {\em Istanbul, 34134, Türkiye}} }
\maketitle

\begin{abstract}
In this paper, employing the exponential corrected entropy (Chatterjee and Ghosh in Phys Rev Lett 125:041302, 2020), we derive the modified Friedmann equations from the first law of thermodynamics at apparent horizon and Verlinde's entropic gravity scenario. First, we derive the modified Friedmann equations from the first law of thermodynamics. We investigate the validity of generalised second law (GSL) of thermodynamics and find that it is always satisfied for the all eras of universe. Moreover, we investigate the deceleration parameter for the case $k=0$ in two frameworks. Finally, we numerically study the bouncing behaviour for the modified Friedmann equations obtained from entropic gravity. The results indicate that the bouncing behaviour is possible for the cases $k=1$ and $k=-1$.
\end{abstract}

\section{Introduction}
\label{intro}
Black hole thermodynamics plays a fundamental role in setting a deep connection among general relativity, thermodynamics, and quantum mechanics \cite{Bekenstein1972,Bekenstein1973,Bardeen1973,Hawking1974,Bekenstein1974,Hawking1975,Carlip2014,Mann2024}. Although this deep connection is not fully known, indeed, it is an outcome of black hole entropy $S$ and temperature $T$ being proportional to horizon area $A_{h}$ and surface gravity $\kappa$, respectively. Drawn inspiration from black hole thermodynamics, Jacobson obtained the Einstein field equations from the first law of black hole thermodynamics \cite{Jacobson1995}. Combining entropy$-$area relation with the Clausius relation, he derived the field equation from thermodynamical viewpoint. Therefore, the field equation can be interpreted as an equation of state. The correspondence between gravity and thermodynamics is generally known as the thermodynamics-gravity conjecture and the research in this direction is rich with the studies in the literature \cite{Padmanabhan2002,Eling2006,Paranjape2006,Kothawala2007,Padmanabhan2007,Cai2005,Akbar2006,Akbar2007,Cai2007a,Cai2007b,Cai2008,Sheykhi2010a,Awad2014,Salah2017,Kouwn2018,Okcu2020,Okcu2024,Luo2023,Alsabbagh2023,Sheykhi2010b,Sheykhi2007a,Sheykhi2007b,Sheykhi2009,Sheykhi2018,Nojiri2019,Lymperis2018,Karami2011,Sheykhi2019,Saridakis2020,Barrow2021,Saridakis2021,Anagnostopoulos2020,Sheykhi2021,Asghari2021,Sheykhi2022,Sheykhi2022b,Asghari2022,Sheykhi2023,Sheykhi2023b,Lymperis2021,Drepanou2022,Odintsov2023,Abreu2022,Coker2023,Ourabah2024,Genarro2022,Odintsov2023b,Nojiri2022,Nojiri2022b,Nojiri2023,Nojiri2023b,Verlinde2011,Cai2010,Shu2010,Sheykhi2010c,Sheykhi2011,Gao2010,Ling2010,FCai2010,Hendi2011,Basilakos2012,Kibaroglu2020,Senay2021,Sefiedgar2017,Feng2018,Abreu2018,Jusufi2023,Jusufi2023b}. For example, obtaining the Einstein field equations  from the first law of thermodynamics is an interesting topic in the thermodynamics-gravity conjecture \cite{Padmanabhan2002,Eling2006,Paranjape2006,Kothawala2007,Padmanabhan2007}. Another interesting topic aims to understand the relation between the first law and Friedmann equations. In Ref. \cite{Cai2005}, defining the first law of thermodynamics as $-dE=TdS$ at the apparent horizon $\tilde{r_{A}}$, Cai and Kim derived the higher dimensional Friedmann equations. Here $-dE$, $T_{h}$ and $dS_{h}$ correspond to the energy flux crossing the horizon, temperature and change in entropy, respectively. Using entropy-area relations of Gauss-Bonnet and Lovelock gravity theories, they obtained the Friedmann equations \cite{Cai2005}. Using the thermodynamical arguments of Ref. \cite{Cai2005}, the Friedmann equations in $f(R)$ gravity and scalar-tensor gravity theories can be found in Ref. \cite{Akbar2006}. Assuming a slow change of the apparent horizon, the temperature is defined by \cite{Cai2005} \footnote{The units $\hbar=c=G_{N}=L^2_{Pl}=k_{B}=1$ are used throughout the paper.} 
\begin{equation}
\label{Temp1}
T_{h}=\frac{1}{2\pi\tilde{r_{A}}}.
\end{equation}
The entropy of apparent horizon is given by \cite{Cai2005}
\begin{equation}
\label{entropy1}
 S_{h}=\frac{A_{h}}{4},
\end{equation}
where the horizon area is given by $A_{h}=4\pi\tilde{r_{A}}^{2}$. The temperature in Eq. (\ref{Temp1}) is not proportional to the surface gravity at the apparent horizon since it is an approximation for the small change of horizon. The surface gravity is defined as follows \cite{Akbar2007,Hayward1998}:
\begin{equation}
\label{kappa}
\kappa=-\frac{1}{\tilde{r_{A}}}\left(1-\frac{\dot{\tilde{r_{A}}}}{2H\tilde{r_{A}}}\right),
\end{equation}
where dot is the derivative with respect to time, and $H$ is the Hubble parameter. From the proportionality of temperature and apparent horizon, we have \cite{Akbar2007}
\begin{equation}
\label{Temp}
T_{h}=\frac{\kappa}{2\pi}=-\frac{1}{2\pi\tilde{r_{A}}}\left(1-\frac{\dot{\tilde{r_{A}}}}{2H\tilde{r_{A}}}\right).
\end{equation}
Taking into account the above equation with the entropy-area relation (\ref{entropy1}) requires the revising the first law of thermodynamics as follows:
\begin{equation}
\label{firstLaw}
dE=T_{h}dS_{h}+WdV,
\end{equation}
where $E$, $W$ and $V$ are total energy inside the apparent horizon, the work density and the volume, respectively. Recently, there has been extensive research in the literature on Friedmann equations and apparent horizon thermodynamics \cite{Cai2007a,Cai2007b,Cai2008,Sheykhi2010a,Awad2014,Salah2017,Kouwn2018,Okcu2020,Okcu2024,Luo2023,Alsabbagh2023,Sheykhi2010b,Sheykhi2007a,Sheykhi2007b,Sheykhi2009,Sheykhi2018,Nojiri2019,Lymperis2018,Karami2011,Sheykhi2019,Saridakis2020,Barrow2021,Saridakis2021,Anagnostopoulos2020,Sheykhi2021,Asghari2021,Sheykhi2022,Sheykhi2022b,Asghari2022,Sheykhi2023,Sheykhi2023b,Lymperis2021,Drepanou2022,Odintsov2023,Abreu2022,Coker2023,Ourabah2024,Genarro2022,Odintsov2023b,Nojiri2022,Nojiri2022b,Nojiri2023,Nojiri2023b}.

The connection between gravity and thermodynamics can also be handled in Verlinde's entropic gravity \cite{Verlinde2011}. Gravity is not regarded as a fundamental force. According to Verlinde, gravity is an entropic force based on two principles, namely, the equipartition law of energy and the holographic principle. When a test particle has a displacement $\Delta x$ from the holographic screen, the entropic force is defined by $F\Delta x=T\Delta S$. Here $\Delta S$ and $T$ correspond to entropy change and temperature of the screen, respectively.  Employing a holographic screen with Unruh temperature yields Newton's second law. Moreover, starting from  the holographic principle and the equipartition law of energy, Newton's gravitational law and Einstein field equations can be derived in entropic gravity scenario. The entropic gravity theory has attracted a lot of attention in recent years. There are many studies devoted to derivations of Newton's gravitational law, Einstein field equations, and Friedmann equations in the literature \cite{Cai2010,Shu2010,Sheykhi2010c,Sheykhi2011,Gao2010,Ling2010,FCai2010,Hendi2011,Basilakos2012,Kibaroglu2020,Senay2021,Sefiedgar2017,Feng2018,Abreu2018,Jusufi2023,Jusufi2023b}. 

Motivated by numerous reasons, modifications of entropy-area relation have been studied in the literature \cite{Govindarajan2001,Mann1998,Sen2013,Das2002,Pourhassan2015,Sadeghi2014,Sadeghi2016,Tsallis2013,Kaniadakis2002,Kaniadakis2005,Barrow2020,Jalalzadeh2021}. Different correction scenarios may provide new insights on both black hole thermodynamics and the evolution of universe. Taking into account quantum gravitational effects at Planck scale yields a well-known logarithmic correction to entropy \cite{Govindarajan2001,Mann1998,Sen2013,Das2002,Pourhassan2015,Sadeghi2014,Sadeghi2016}. For example, loop quantum gravity \cite{Cai2008,Sheykhi2010a}, generalized uncertainty principle (GUP) \cite{Awad2014,Salah2017,Okcu2020,Okcu2024}, extended uncertainty principle (EUP) \cite{Luo2023,Kouwn2018}, generalized and extended uncertainty principle (GEUP) \cite{Alsabbagh2023} and rainbow gravity \cite{Sefiedgar2017,Feng2018} modifications of Friedmann equations can be found in the literature.  The modifications of standard uncertainty principle imply a non singular beginning of universe due to the minimal length notion \cite{Awad2014,Salah2017,Okcu2020,Okcu2024,Alsabbagh2023}. Another interesting modification of entropy-area relation is based on the generalized statistics. For example, Kaniadakis statistics originates from the relativistic statistics mechanics \cite{Kaniadakis2002,Kaniadakis2005}. Kaniadakis entropy modified Friedmann equations were investigated in Refs. \cite{Sheykhi2023b,Lymperis2021,Drepanou2022,Abreu2018}. Apart from logarithmic correction, there is another form of correction suggesting that the entropy is a power law function of its area \cite{Tsallis2013,Barrow2020,Jalalzadeh2021}. For example, Tsallis statistics suggests that entropy must be nonadditive \cite{Tsallis2013}. Tsallis modified Friedmann equations can be found in Refs. \cite{Sheykhi2018,Nojiri2019,Lymperis2018}. Inspired by COVID-19, Barrow proposed that the black hole area may be deformed due to the quantum gravitational effects \cite{Barrow2020}. The extensions of Barrow entropy to cosmological applications can be found in Refs. \cite{Saridakis2020,Barrow2021,Saridakis2021,Anagnostopoulos2020,Sheykhi2021,Asghari2021,Sheykhi2022}. Just like Tsallis and Barrow entropies, fractional entropy based on fractional quantum mechanics also suggests that entropy is a power law function of its area \cite{Jalalzadeh2021}. In Ref. \cite{Coker2023}, we obtained the fractional entropy modified Friedmann equations from first law and entropic gravity.

The leading order correction to entropy-area relation may be regarded as logarithmic \cite{Govindarajan2001,Mann1998,Sen2013,Das2002,Pourhassan2015,Sadeghi2014,Sadeghi2016}. The logarithmic term becomes important when the black hole size shrinks. For example, thermal fluctuations around equilibrium temperature lead to a logarithmic corrected entropy-area relation \cite{Das2002}. This correction can be considered  a perturbative correction. However, non-perturbative corrections to black hole entropy are inevitable when black hole becomes even smaller due to the Hawking radiation. Recently, authors of Ref. \cite{Chatterjee2020} argued that the black hole entropy has such a correction as an exponential term. They obtained exponential correction to entropy only employing horizon geometry without resorting to loop quantum gravity or string theory. Counting only the microstates for quantum states residing on black hole horizon may yield an exponential correction in black hole entropy. The exponential corrected entropy is given by \cite{Chatterjee2020}
\begin{equation}
\label{exponentialEntropy}
S=\frac{A}{4}+\alpha e^{-A/4}.
\end{equation}
In order to track the effects of exponential correction, we add the parameter $\alpha$. This idea can originally be traced back to Ref. \cite{Pourhassan2015}.  In the limit $\alpha\rightarrow0$, the above expression reduces to semiclassical result, i.e., Eq. (\ref{entropy1}).  The exponential correction is negligible for large areas, while it gives a remarkable contribution to black hole entropy when black hole becomes small. The effects of exponential correction were investigated for black hole thermodynamics in Refs. \cite{Pourhassan2021,Pourhassan2021b}. However, the effects of exponential entropy have not yet been studied for cosmology. The non-perturbative modification as an exponential term generates more sensitive correction to Friedmann equations at small length scales than the perturbative correction. Therefore, in this paper, we consider the modified Friedmann equations for exponential corrected entropy. In order to derive the modified Friedmann equations, we use two different approaches, i.e., the first law of thermodynamics at apparent horizon and entropic gravity case. We try to understand the effects of exponential correction for the cosmology.

The paper is organised as follows: In Section \ref{sec2}, we derive the modified Friedmann equations from the first law of thermodynamics at apparent horizon. In order to grasp the effects of exponential correction, we compute the deceleration parameter. Then, we investigate the validity of GSL. In Section \ref{sec3}, we obtain the modified Friedmann equations from entropic gravity scenario. Similarly, we compute the deceleration parameter. Then, we investigate the bouncing behaviour of modified Friedmann equations in entropic gravity case. Finally, we close this paper in Section \ref{sec4}.

\section{Friedmann equations from the first law of thermodynamics}
\label{sec2}
First, we begin to review the basic elements of Friedmann-Robertson-Walker (FRW) universe. In the compact form, the line element is given by
\begin{equation}
 \label{lineElement}
 ds^{2}=h_{ab}dx^{a}dx^{b}+\tilde{r}^{2}d\Omega^{2},
 \end{equation}
where\, \, $\tilde{r} \,=\, a(t)r$, \, \,the coordinates $x^a \,=\, (t,r)$\,\, \,and \, two-dimensional \, metric  \, \,$ h_{ab}=\\ diag\left(-1,a^{2}/(1-kr^{2})\right)$. Here the spatial curvature constant $k=$ $-1$, $0$ and $1$ correspond to open, flat and closed universe, respectively. The apparent horizon $\tilde{r_{A}}$ is defined by
 \begin{equation}
 \label{apparentHor}
 \tilde{r_{A}}=ar=\frac{1}{\sqrt{H^{2}+k/a^{2}}},
 \end{equation}
where $H=\dot{a}/a$. We assume the energy and matter of universe as a perfect fluid. So we can write energy-momentum tensor as follows:
 \begin{equation}
 \label{energyMomentumTensor}
 T_{\mu\nu}=(\rho+p)u_{\mu}u_{\nu}+pg_{\mu\nu}.
 \end{equation}
Here $\rho$, $p$ and $u^{\mu}$ correspond to energy density, pressure and four-velocity of fluid, respectively. From the conservation of energy-momentum tensor, i.e., $\nabla_{\mu}T^{\mu\nu}=0$, one can obtain the continuity equation as follows:
 \begin{equation}
 \label{continuityEqu}
 \dot{\rho}+3H(\rho+p)=0.
 \end{equation}
The work density $W$ is given by \cite{Hayward1998}
  \begin{equation}
 \label{workDensity}
 W=-\frac{1}{2}T^{ab}h_{ab}=\frac{1}{2}(\rho-p).
 \end{equation}
Here, work density $W$ is the the work by the volume change of the universe. The volume is given by \cite{Akbar2007}
 \begin{equation}
 V=\frac{4}{3}\pi \tilde{r_{A}}^{3}.
\label{volume}
\end{equation}
Let us begin to calculate the terms of first law in Eq. (\ref{firstLaw}). Using the Eqs. (\ref{continuityEqu}) and (\ref{volume}), $dE$ is given by
\begin{equation}
 \label{dE}
 dE=\rho dV+Vd\rho=4\pi\rho\tilde{r_{A}}^{2}d\tilde{r_{A}}-4\pi(\rho+p)\tilde{r_{A}}^{3}Hdt.
 \end{equation}
From Eqs. (\ref{workDensity}) and (\ref{volume}), the term $WdV$  is found
 \begin{equation}
\label{WdV}
WdV=2\pi(\rho-p)\tilde{r_{A}}^{2}d\tilde{r_{A}}.
\end{equation}
The general expression of entropy is given by \cite{Awad2014}
\begin{equation}
\label{f(A)Rel}
S_{h}=\frac{f(A_{h})}{4},
\end{equation}
and its differential is expressed 
\begin{equation}
\label{diffOfEntropy}
\frac{dS_{h}}{dA_{h}}=\frac{f'(A_{h})}{4},
\end{equation}
where prime is the derivative with respect to area. From Eq. (\ref{exponentialEntropy}) and the above expression, one can find
\begin{equation}
\label{diffOfEntropy2}
f'(A_{h})=1-\alpha e^{-\pi \Tilde{r_{A}}^{2}} .
\end{equation}
At last, one can obtain the term $T_{h}dS_{h}$ as
\begin{equation}
\label{TdS}
T_{h}dS_{h}=-\left(1-\frac{\dot{\tilde{r_{A}}}}{2H\tilde{r_{A}}}\right)f'(A_{h})d\tilde{r_{A}}.
\end{equation}
Combining the terms $dE$,$WdV$, $T_{h}dS_{h}$ in the first law of thermodynamics at apparent horizon and employing the differential of apparent horizon
\begin{equation}
\label{difAppa}
d\tilde{r_{A}}=-H\tilde{r_{A}}^{3}\left(\dot{H}-\frac{k}{a^{2}}\right)dt,
\end{equation}
we obtain
\begin{equation}
\label{dynamicalEq}
4\pi(\rho+p)\tilde{r_{A}}^{3}Hdt=f'(A_{h})d\tilde{r_{A}}.
\end{equation}
Employing the continuity equation (\ref{continuityEqu}) in Eq. (\ref{dynamicalEq}), one can obtain the differential form of Friedmann equation as follows:
\begin{equation}
 \label{firstEquation}
\frac{f'(A_{h})}{\tilde{r_{A}}^{3}}d\tilde{r_{A}}=-\frac{4\pi}{3}d\rho.
\end{equation}
Using Eq. (\ref{diffOfEntropy2}) in Eq. (\ref{firstEquation}) and integrating the above expression, we obtain the first Friedmann equation
\begin{equation}
\label{firstFriedmannEquationa}
-\frac{1}{2\tilde{r_{A}}^{2}}+\frac{\alpha}{2\tilde{r_{A}}^{2}}\left[e^{-\pi\tilde{r_{A}}^{2}}+\pi\tilde{r_{A}}^{2}Ei\left(-\pi\tilde{r_{A}}^{2}\right)\right]=-\frac{4\pi}{3}\rho,
\end{equation}
where we set the integration constant to zero. Here $Ei$ is the exponential integral function. The correction term gives a remarkable contribution to Friedmann equation in small length scale. However, the above equation does not imply a non-zero minimum apparent horizon. Instead, the apparent horizon is still allowed to go to zero. Thus, the modified Friedmann equation does not remove the initial singularity. From Eqs. (\ref{diffOfEntropy2}), (\ref{difAppa}) and (\ref{dynamicalEq}), the second Friedmann equation is given by
\begin{equation}
\label{secondEquation}
-4\pi(\rho+p)=\left(\dot{H}-\frac{k}{a^{2}}\right)(1-\alpha e^{-\pi\tilde{r_{A}}^{2}}).
\end{equation}
In the limit $\alpha\rightarrow0$, these modified Friedmann equations reduce to standard Friedmann equations. Using Eq. (\ref{apparentHor}), the modified Friedmann equations can be written in term of Hubble parameter \footnote{The exponential integral correction term to first Friedmann equation was also reported in Ref. \cite{Ourabah2024}. Recently, using modified gravitational potentials in Verlinde's entropic gravity scenario, various entropies were constructed in Ref. \cite{Ourabah2024}.  The author obtained the following entropy
\begin{equation}
\label{YukawaEnt}
S=\frac{1}{4}\left[A-2\beta\lambda^{2}e^{-\sqrt{\frac{A}{4\pi\lambda^{2}}}}\left(12\pi+6\sqrt{\frac{\pi A}{\lambda^{2}}}+\frac{A}{\lambda^{2}}\right)\right]
\end{equation}
from Yukawa-type of potential (See Eq. (6) in Ref. \cite{Ourabah2024}). Here, $\beta$ and $\lambda$ correspond to strength and range, respectively. Employing the the above entropy in the first law of thermodynamics at apparent horizon, the author obtained the following Friedmann equation as follows:
\begin{equation}
\label{YukawaFriedmann}
\left(H^{2}+\frac{k}{a^{2}}\right)-\frac{\beta}{\lambda^{2}}\left[\lambda\left(H^{2}+\frac{k}{a^{2}}\right)e^{\frac{1}{\lambda\left(\sqrt{H^{2}+k/a^{2}}\right)}}\left(-\lambda+\frac{\left(1-2\sqrt{\pi}\right)}{\sqrt{H^{2}+k/a^{2}}}\right)+\left(1-2\sqrt{\pi}\right)Ei\left(\frac{-1}{\lambda\sqrt{H^{2}+k/a^{2}}}\right)\right]=\frac{8\pi\rho}{3}.
\end{equation}
}
\begin{align}
\label{Fried2}
&\left(H^{2}+\frac{k}{a^{2}}\right)\left[1-\alpha\left(e^{-\pi\left(H^{2}+\frac{k}{a^{2}}\right)^{-1}}+\pi\left(H^{2}+\frac{k}{a^{2}}\right)^{-1}Ei\left(-\pi\left(H^{2}+\frac{k}{a^{2}}\right)^{-1}\right)\right)\right]=\frac{8\pi\rho}{3},\nonumber\\
&\left(\dot{H}-\frac{k}{a^{2}}\right)\left(1-\alpha e^{-\pi\left(H^{2}+\frac{k}{a^{2}}\right)^{-1}}\right)=-4\pi(\rho+p).
\end{align}
Now, in order to see the effects of exponential correction, we consider the deceleration parameter. The deceleration parameter $q$ is defined by
\begin{equation}
\label{decelParam}
q=1-\frac{\dot{H}}{H^{2}},
\end{equation}
where the case $q>0$ corresponds to decelerated expansion phase of universe, while the case $q<0$ corresponds to accelerated expansion phase of universe. Using the modified Friedmann equations (\ref{Fried2}) and choosing equation of state as $p=\omega \rho$, one can obtain the expression for the declaration parameter as
\begin{equation}
\label{decelParam2}
q=-1+\frac{3}{2}\frac{(1+\omega)}{1-\alpha e^{-\pi/H^{2}}}\left[1-\alpha\left(e^{-\pi/H^{2}}+\frac{\pi}{H^{2}}Ei\left(-\frac{\pi}{H^{2}}\right)\right)\right].
\end{equation}
Here we consider the flat case $k=0$ in agreement with cosmological observations \cite{Odintsov2011}. Since quantum gravity effects are sufficiently tiny for the radiation and matter dominated eras, one can expand the deceleration parameter
\begin{equation}
\label{decelSeries}
q=-1+\frac{3(1+\omega)}{2}-\frac{3\pi\alpha(1+\omega)}{2H^{2}}Ei\left(-\frac{\pi}{H^{2}}\right)+... \,.
\end{equation}
The deceleration parameter is given by
\begin{equation}
\label{qMattAndRad}
q_{m}=\frac{1}{2}-\frac{3\pi\alpha}{2H^{2}}Ei\left(-\frac{\pi}{H^{2}}\right),\qquad\qquad\qquad\qquad q_{r}=1-\frac{2\pi\alpha}{H^{2}}Ei\left(-\frac{\pi}{H^{2}}\right)
\end{equation}
for the matter ($\omega=0$) and radiation ($\omega=1/3$) dominated eras, respectively. For $H>0$, the correction terms give positive contribution to deceleration parameter. Therefore, the universe is more decelerated for the matter and radiation dominated eras. 

\subsection{Generalised second law of thermodynamics}
\label{sec2.1}

Now, we consider the total entropy change of universe in the presence of exponential correction. Reorganising Eq. (\ref{dynamicalEq}), we obtain
\begin{equation}
\label{raWRTT}
\dot{\tilde{r_{A}}}=\frac{4\pi(\rho+p)\tilde{r_{A}}^{3}H}{1-\alpha e^{-\pi\tilde{r_{A}}^{2}}}
\end{equation}
Substituting the above equation and Eq. (\ref{diffOfEntropy2}) into Eq. (\ref{TdS}), we find
\begin{equation}
\label{entWRTT}
T_{h}\dot{S_{h}}=4\pi(\rho+p)H\tilde{r_{A}}^{3}\left[1-\frac{2\pi(\rho+p)\tilde{r_{A}}^{2}}{1-\alpha e^{-\pi\tilde{r_{A}}^{2}}}\right].
\end{equation}
The above expression does not guarantee the validity of second law. The second law of thermodynamics may be violated since $(\rho+p)$ is negative during the accelerated phase. Therefore, we need the consider matter fields inside the horizon. Thus, we must check the validity of GSL. Let us begin to consider the Gibbs equation given by \cite{Izquierdo2006}
\begin{equation}
\label{GibbsEqu}
T_{f}dS_{f}=d(\rho V)+pdV=Vd\rho+(p+\rho)dV,
\end{equation}
where $S_{f}$ and $T_{f}$ are the entropy and temperature of matter fields inside the horizon. We assume the thermal equilibrium between the apparent horizon and matter fields, i.e., $T_{h}=T_{f}$. This assumption prevents the energy flows between the horizon and matter fields. Thus, the FRW geometry is preserved \cite{Izquierdo2006}. From Eqs. (\ref{continuityEqu}), (\ref{volume}), (\ref{raWRTT}) and Gibbs equation (\ref{GibbsEqu}), the entropy change of matter fields inside the apparent horizon is given by
\begin{equation}
\label{flEntWRTT}
T_{h}\dot{S_{f}}=-4\pi(\rho+p)H\tilde{r_{A}}^{3}\left(1-\frac{4\pi(\rho+p)\tilde{r_{A}}^{2}}{1-\alpha e^{-\pi\tilde{r_{A}}^{2}}}\right).
\end{equation}
Finally, from Eq. (\ref{entWRTT}) and the above equation, the total entropy evolution of universe is expressed as follows:
\begin{equation}
\label{GSL}
T_{h}(\dot{S_{h}}+\dot{S_{f}})=\frac{8\pi^{2}(\rho+p)^{2}H\tilde{r_{A}}^{5}}{1-\alpha e^{-\pi\tilde{r_{A}}^{2}}}.
\end{equation}
In the limit $\alpha\rightarrow0$, it is clear that the above expression reduces to the standard form and GSL is always satisfied. It may seem that $1-\alpha e^{-\pi\tilde{r_{A}}^{2}}> 0$ must hold to satisfy the GSL. This condition yields a lower bound $\tilde{r_A}>\sqrt{\frac{\ln\alpha}{\pi}}$. However, this lower bound is imaginary in the range $\alpha\in(0,1)$. For $\alpha=1$,  the condition imply that $\tilde{r_A}>0$. The condition  $1-\alpha e^{-\pi\tilde{r_{A}}^{2}}> 0$ always holds for $\tilde{r_A}>0$. Thus, the right hand side of Eq. (\ref{GSL}) never decreases with respect to time. Therefore, the GSL is always valid during the all eras of the universe for any spatial curvature.

\section{Friedmann equations from the entropic gravity}
\label{sec3}
Based on the arguments of Refs. \cite{Verlinde2011,Cai2010}, we are going to obtain the modified Friedmann equations. First, we consider a compact spatial region $\mathcal{V}$ with a holographic screen on the boundary $\mathcal{\partial V}$. Considering the number of the bits on the holographic screen 
\begin{equation}
\label{NARel}
N=A_{h},
\end{equation}
with Eq. (\ref{entropy1}), the relation between $N$ and $S_{h}$ is given by
\begin{equation}
N=4S_{h}.
\label{NSRelation}
\end{equation}
Another important assumption is that the total entropy of the screen is given by the equipartition law of energy. It is defined by
\begin{equation}
\label{equPart}
E=\frac{1}{2}NT_{U},
\end{equation}
where $T_{U}$ is the Unruh temperature of the screen. The Unruh temperature is given by 
\begin{equation}
\label{unTemp}
T=\frac{a_{r}}{2\pi}=-\frac{\ddot{a}r}{2\pi},
\end{equation}
where $a_{r}$ corresponds to the acceleration. It is defined by \cite{Cai2010}
\begin{equation}
\label{accl}
a_{r}=-\frac{d^{2}\tilde{r_{A}}}{dt^{2}}=-\ddot{a}r.
\end{equation}
We also consider the active gravitational mass $\mathcal{M}$ instead of total mass $M$ inside the horizon. $\mathcal{M}$ is given by \cite{Cai2010}
\begin{equation}
\label{KomarMss}
\mathcal{M}=2\int_{\mathcal{V}}dV\left(T_{\mu\nu}-\frac{1}{2}Tg_{\mu\nu}\right)u^{\mu}u^{\nu}=\frac{4}{3}\pi(\rho+3p)\widetilde{r_{A}}^{3}.
\end{equation}
Using Eqs. (\ref{exponentialEntropy}) and (\ref{NSRelation}), the modified number of bits is given by
\begin{equation}
\label{ModifiedN}
N=4\pi\tilde{r_{A}}^{2}+4\alpha e^{-\pi\tilde{r_{A}}^{2}}.
\end{equation}
Considering the assumption $\mathcal{M}=E$, from Eqs. (\ref{equPart})-(\ref{ModifiedN}), we can derive the acceleration equation
\begin{equation}
\label{EntropicDynamicEq}
-\frac{\ddot{a}}{a}\left(1+\frac{\alpha}{\pi\tilde{r_{A}}^{2}}e^{-\pi\tilde{r_{A}}^{2}}\right)=\frac{4\pi}{3}(\rho+3p).
\end{equation}
Multiplying $2a\dot{a}$ on the both side of the above equation and employing the continuity equation (\ref{continuityEqu}), we find
\begin{equation}
\label{integralFirstEq}
\int\frac{d\dot{a}^{2}}{dt}dt=\frac{8\pi}{3}\int\frac{1}{\left(1+\frac{\alpha e^{-\pi a^{2}r^{2}}}{\pi a^{2}r^{2}}\right)}\frac{d(\rho a^{2})}{dt}dt.
\end{equation}
Moreover, in order to solve this integral, we consider the equation of state as $p=\omega\rho$. Using this equation of state in continuity equation (\ref{continuityEqu}) yields
\begin{equation}
\label{solutionContiEq}
\rho=\rho_{0}a^{-3(1+\omega)},
\end{equation}
where $\rho_{0}$ is a constant. From the above equation, we get
\begin{equation}
\label{dRhoa2}
d(\rho a^{2})=-\rho_{0}(1+3\omega)a^{-2-3\omega}da.
\end{equation}
Using Eq. (\ref{dRhoa2}) and equation of state $p=\omega\rho$ in Eq. (\ref{integralFirstEq}), the integral is given by
\begin{equation}
\label{integralFirstEq2}
\int\frac{d\dot{a}^{2}}{dt}dt=-\frac{8\pi\rho_{0}(1+\omega)}{3}\int\frac{a^{-2-3\omega}}{\left(1+\frac{\alpha e^{-\pi a^{2}r^{2}}}{\pi a^{2}r^{2}}\right)}\frac{da}{dt}dt.
\end{equation}
In order to solve this integral, we consider the small exponential correction. Then, the above equation can be written by 
\begin{equation}
\label{integrakFirstEq3}
\int\frac{d\dot{a}^{2}}{dt}dt=-\frac{8\pi\rho_{0}(1+\omega)}{3}\int a^{-2-3\omega}\left(1+\frac{\alpha e^{-\pi a^{2}r^{2}}}{\pi a^{2}r^{2}}\right)\frac{da}{dt}dt,
\end{equation}
where we use the assumption $(1+\beta)^n\approx1+\beta n$ for $\beta\ll 1$. Solving this integral, one can obtain the first Friedmann equation as follows:
\begin{equation}
\label{EntropicFirstEq}
\left(\frac{\dot{a}}{a}\right)^{2}+\frac{k}{a^{2}}=\frac{8\pi\rho}{3}\left(1-\frac{\alpha(3\omega+1)}{2\pi\tilde{r_{A}}^{2}}E_{\frac{3\omega+5}{2}}(\pi\tilde{r_{A}}^{2})\right),
\end{equation}
where $E_{n}$ is the exponential integral function. Using $H=\dot{a}/a$, the Friedmann equations are written in term of Hubble parameter
 \begin{eqnarray}
 \label{FriedmanEntropH}
 H^{2}+\frac{k}{a^{2}}=\frac{8\pi\rho}{3}\left[1-\frac{\alpha(3\omega+1)}{2\pi}\left(H^{2}+\frac{k}{a^{2}}\right)E_{\frac{3\omega+5}{2}}\left(\pi\left(H^{2}+\frac{k}{a^{2}}\right)^{-1}\right)\right],\nonumber\\
 (\dot{H}+H^{2})\left[1+\frac{\alpha}{\pi}\left(H^{2}+\frac{k}{a^{2}}\right)e^{-\pi\left(H^{2}+\frac{k}{a^{2}}\right)^{-1}}\right]=-\frac{4\pi}{3}(\rho+3p).
 \end{eqnarray}
As can be seen, the Friedmann equations obtained in this section are different from the Friedmann equations obtained in the previous section. In the previous section, the modified Friedmann equations are exactly derived. In this section, we approximately obtain the modified Friedmann equations. Moreover, the Friedmann equations obtained in this section depend on a specific equation of state, i.e., $p=\omega \rho$. Therefore, Eqs. (\ref{Fried2}) and (\ref{FriedmanEntropH}) are not the same.
 
 Again, we consider the deceleration parameter for $k=0$. From Eqs. (\ref{decelParam}) and (\ref{FriedmanEntropH}), the deceleration parameter is given by
 \begin{equation}
\label{decelEntropic}
q=\frac{3\omega+1}{2}\left[1-\frac{\alpha(3\omega+1)H^{2}}{2\pi}E_{\frac{3\omega+5}{2}}\left(\frac{\pi}{H^{2}}\right)\right]^{-1}\left(1+\frac{\alpha}{\pi}H^{2}e^{-\frac{\pi}{H^{2}}}\right)^{-1}.
 \end{equation}
 For small quantum gravity effects, the above equation can  be approximated as
 \begin{equation}
\label{decelEntropicSeries}
q\approx\frac{3\omega+1}{2}\left[1+\frac{\alpha e^{-\pi/H^{2}}H^{2}}{2\pi}\left((1+3\omega)e^{\pi/H^{2}}E_{\frac{3\omega+5}{2}}\left(\frac{\pi}{H^{2}}\right)-2\right)\right].
 \end{equation}
The deceleration parameter is given by
\begin{equation}
\label{decelEntropicMatAndRad}
q_{m}=\frac{1}{2}\left[1-\frac{\alpha e^{-\pi/H^{2}}H^{2}}{2\pi}\left(2-e^{\pi/H^{2}}E_{\frac{5}{2}}\left(\frac{\pi}{H^{2}}\right)\right)\right],\qquad q_{r}=1-\frac{\alpha e^{-\pi/H^{2}}H^{2}}{\pi}\left(1-e^{\pi/H^{2}}E_{3}\left(\frac{\pi}{H^{2}}\right)\right)
\end{equation}
for the matter ($\omega=0$) and radiation ($\omega=1/3$) dominated eras, respectively.
The modified deceleration parameter of radiation dominated era is positive for the condition
\begin{equation}
\label{rCondition}
\frac{\pi}{e^{-\pi/H^{2}}H^{2}\left(1-e^{\pi/H^{2}}E_{3}\left(\frac{\pi}{H^{2}}\right)\right)}>\alpha.
\end{equation}
For the matter dominated era, deceleration parameter is positive when the condition
\begin{equation}
\label{mCondition}
\frac{2\pi e^{\pi/H^{2}}}{H^{2}\left(2-e^{\pi/H^{2}}E_{5/2}\left(\frac{\pi}{H^{2}}\right)\right)}>\alpha
\end{equation}
is satisfied.

\subsection{Bouncing universe}
\label{sec3.1}
In this section, we are going to investigate the bouncing behaviour for the modified Friedmann equations obtained from Verlinde's entropic gravity. The most striking feature of bouncing behaviour is the absence of the initial singularity. In the bouncing scenario, the universe begins from a contraction phase with $\dot{a}<0$, goes through a non-vanishing and minimal scale factor $a_{c}$ (critical point), then starts to accelerate in expansion phase. The bouncing behaviour has been intensively studied in the literature \cite{Jusufi2023,Ashtekar2006,Bojowald2001,Ling2009,Ling2010b,Bozza2009,Cai2007,Cai2009Bou,Cai2009Boub,Cai2009Bouc,Cai2009Boud,Cai2010Bou,Cai2011Bou,Cai2012Bou,Corichi2011,Vakili2014,Pan2016,Agrawal2021,Agrawal2023,Agrawal2023b}. Mathematically, the bouncing conditions are expressed as $a_{c}>0$, $\ddot{a_{c}}>0$, $H|_{a=a_{c}}=0$ at the critical point $a_{c}$ \cite{Ling2009,Pan2016}.

The correction terms in the first Friedmann equation (\ref{FriedmanEntropH}) can be written by
\begin{equation}
\label{simpleFriedmann}
H^{2}+\frac{k}{a^{2}}=\frac{8\pi\rho}{3}\left[1-\frac{4\alpha(3\omega+1)\rho}{3}E_{\frac{3\omega+5}{2}}\left(\frac{3}{8\rho}\right)\right].
\end{equation}
At the critical point, the above equation is given by
\begin{equation}
\label{bouncingFriedman1}
\frac{k}{a_{c}^{2}}=\frac{8\pi\rho_{c}}{3}\left[1-\frac{4\alpha(3\omega+1)\rho_{c}}{3}E_{\frac{3\omega+5}{2}}\left(\frac{3}{8\rho_{c}}\right)\right],
\end{equation}
where $\rho_{c}$ is the energy density at the critical point. Solving this equation, we find the solution of the critical scale factor
\begin{equation}
\label{criticalScaleFactor}
a_{c}=\frac{\pm3\sqrt{k}}{2\sqrt{2\pi\rho_{c}\left[3-4\alpha\left(1+3\omega\right)\rho_{c}E_{\frac{3\omega+5}{2}}\left(\frac{3}{8\rho_{c}}\right)\right]}}.
\end{equation}
We consider the positive solution since it is only physically meaningful. At the critical point, the acceleration equation (\ref{EntropicDynamicEq}) is represented by
 \begin{equation}
\label{dynamicalEqBounc}
\frac{\ddot{a_{c}}}{a_{c}}=-\frac{4\pi}{3}\left(\rho_{c}+3p_{c}\right)\left[1-\frac{8\alpha\rho_{c}}{3}e^{-\frac{3}{8\rho_{c}}}\right],
 \end{equation}
where $p_{c}$ is the pressure at critical point.

The condition $\ddot{a_{c}}>0$ is satisfied for two cases. The first case corresponds to $\left(\rho_{c}+3p_{c}\right)>0$ and $1-\frac{8\alpha\rho_{c}}{3}e^{-\frac{3}{8\rho_{c}}}<0$, while the second case corresponds to $\left(\rho_{c}+3p_{c}\right)<0$ and $1-\frac{8\alpha\rho_{c}}{3}e^{-\frac{3}{8\rho_{c}}}>0$. Here we consider the first case. Using the equation of state $p=\omega\rho$ and Eq. (\ref{criticalScaleFactor}), Eq. (\ref{dynamicalEqBounc}) is expressed by
\begin{equation}
\label{dynamicalEqBounc2}
\ddot{a_{c}}=\frac{-\sqrt{2\pi k}\left(1+3\omega\right)\rho_{c}}{\sqrt{\rho_{c}\left[3-4\alpha\left(1+3\omega\right)\rho_{c}E_{\frac{3\omega+5}{2}}\left(\frac{3}{8\rho_{c}}\right)\right]}}\left[1-\frac{8\alpha\rho_{c}}{3}e^{-\frac{3}{8\rho_{c}}}\right].
\end{equation}
We consider the condition $(3\omega+1)>0$. In order to satisfy $\ddot{a_{c}}>0$, the condition  $1-\frac{8\alpha\rho_{c}}{3}e^{-\frac{3}{8\rho_{c}}}<0$ yields 
\begin{equation}
\label{aCPosCondRho}
\rho_{c}>\frac{3}{8W(\alpha)},
\end{equation}
where $W$ is the Lambert function.

In order to get $a_{c}>0$ and $\ddot{a_{c}}>0$, we impose the condition
\begin{equation}
\label{condtForAC}
3-4\alpha\left(1+3\omega\right)\rho_{c}E_{\frac{3\omega+5}{2}}\left(\frac{3}{8\rho_{c}}\right)>0,
\end{equation}
from Eqs. (\ref{criticalScaleFactor}) and (\ref{dynamicalEqBounc2}) for the case $k=1$. Taking $\alpha=1$, the $a_{c}$ is positive in the range $0<\rho_{c}<1.753$ from the above condition for the matter dominated universe $\omega=0$. From the conditions in Eqs. (\ref{aCPosCondRho}) and (\ref{condtForAC}), we find that $\ddot{a_{c}}$ is positive in the range $0.661<\rho_{c}<1.753$. Thus, we conclude that the bouncing behaviour is available in the range $0.661<\rho_{c}<1.753$ since $a_{c}$ and $\ddot{a_{c}}$ are positive.

\begin{figure}
\centering
\includegraphics[width=8.18cm]{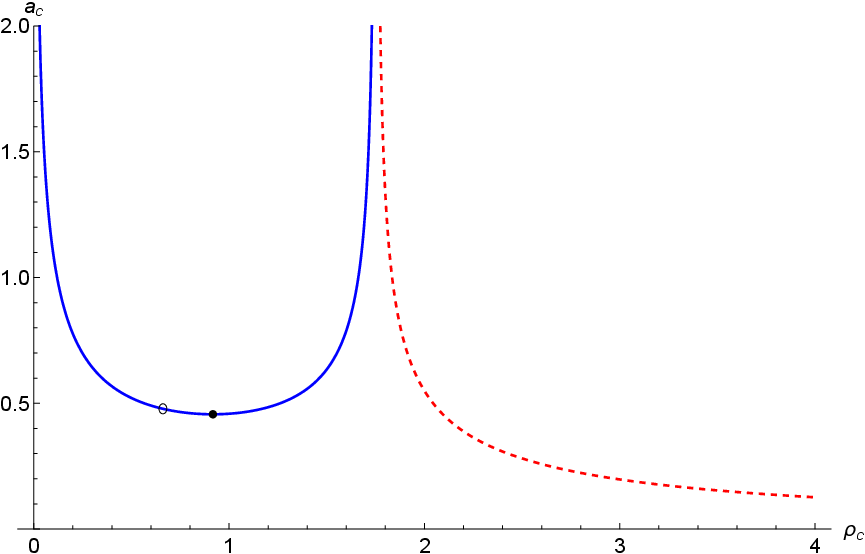}
\includegraphics[width=8.18cm]{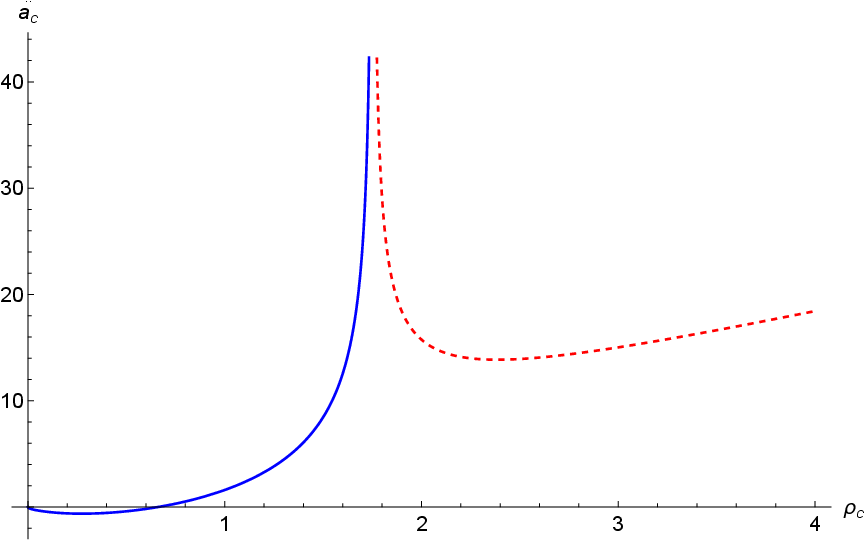}
\caption{(a) Critical scale factor vs critical energy density. (b) Critical $\ddot{a_{c}}$ vs critical energy density. Blue solid and red dashed lines correspond to closed ($k=1$) and open ($k=-1$) cases, respectively. The empty circle shows the value of $\rho_{c}$ where $\ddot{a_{c}}$ vanishes, while the black circle shows the value of $\rho_{c}$ where $a_{c}$ is minimum. We take $\alpha =1$ and $\omega=0$.}
\label{dustScaleVsRho}%
\end{figure}

Let us now consider the radiation dominated universe $\omega=1/3$. This case is relevant to the early stage of the Universe since radiation dominated. Taking $\alpha=1$, we find that $a_{c}$ is positive in the range $0<\rho_{c}<1.25$ from Eq. (\ref{condtForAC}). Using Eqs. (\ref{aCPosCondRho}) and (\ref{condtForAC}), we find that $\ddot{a_{c}}$ is positive in the range $0.661<\rho_{c}<1.25$. We see that the bouncing behaviour is available in the range $0.661<\rho_{c}<1.25$. 

Now, we consider the case $k=-1$. From Eqs. (\ref{criticalScaleFactor}) and (\ref{dynamicalEqBounc2}), we impose the condition
\begin{equation}
\label{condtForAC2}
3-4\alpha\left(1+3\omega\right)\rho_{c}E_{\frac{3\omega+5}{2}}\left(\frac{3}{8\rho_{c}}\right)<0.
\end{equation}
For the matter and radiation dominated universes, taking $\alpha=1$, we obtain $a_{c}>0$ and $\ddot{a_{c}}>0$ in the ranges $\rho_{c}>1.753$ and $\rho_{c}>1.25$, respectively. Therefore, we conclude that the bouncing behaviour is possible for the case $k=-1$.

\begin{figure}
\centering
\includegraphics[width=8.18cm]{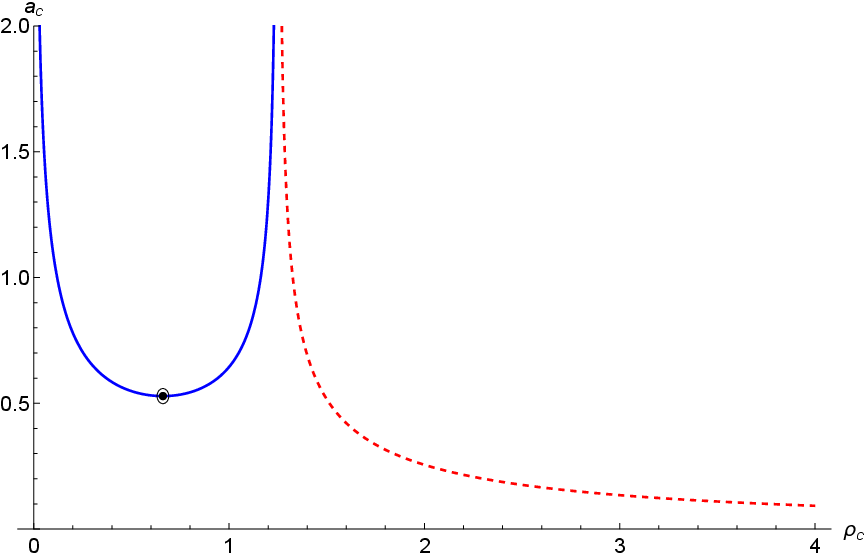}
\includegraphics[width=8.18cm]{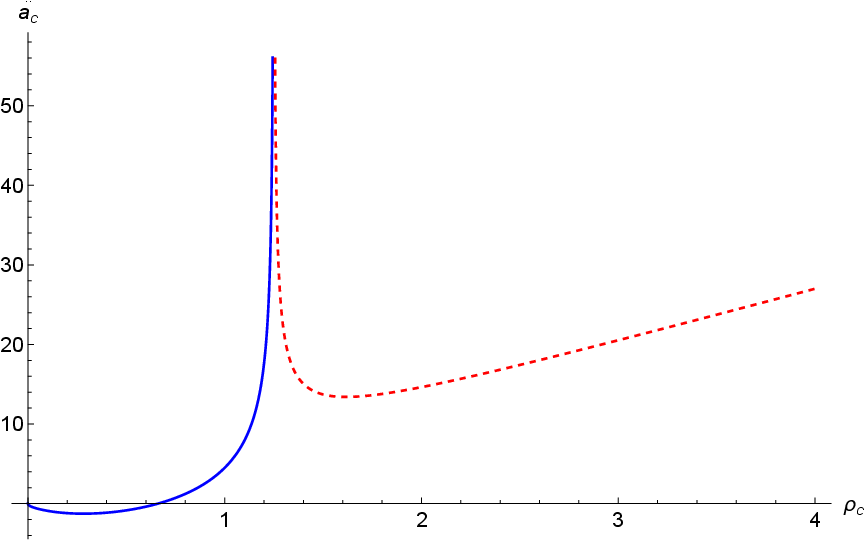}
\caption{(a) Critical scale factor vs critical energy density. (b) Critical $\ddot{a_{c}}$ vs critical energy density. Blue solid and red dashed lines correspond to closed ($k=1$) and open ($k=-1$) cases, respectively. The empty circle shows the value of $\rho_{c}$ where $\ddot{a_{c}}$ vanishes, while the black circle shows the value of $\rho_{c}$ where $a_{c}$ is minimum. We take $\alpha =1$ and $\omega=1/3$.}
\label{radScaleVsRho}%
\end{figure}

In Figs. \ref{dustScaleVsRho} and \ref{radScaleVsRho}, $a_{c}$ and $\ddot{a_{c}}$ are plotted with respect to $\rho_{c}$ for matter and radiation cases, respectively. As can be seen from the figures, $a_{c}$ and $\ddot{a_{c}}$ exhibit different behaviours for closed ($k=1$) and open ($k=-1$) universes. For the case $k=1$ (blue line), $a_{c}$ has a minimum value at $\rho_{c}=0.917$ and $\rho_{c}=0.661$ for matter and radiation cases, respectively. The minimum $a_{c}$ is $0.456$ and $0.529$ for matter and radiation cases, respectively. In Figs. \ref{dustScaleVsRho}.a and \ref{radScaleVsRho}.a, the minimum $a_{c}$ is indicated by a black circle. The minimum $a_{c}$ separates the curve into two branches. $a_{c}$ decreases with respect to $\rho_{c}$ in the left branch, while it increases  with respect to $\rho_{c}$ in the right branch. Although $a_{c}$ is positive in two branches, $\ddot{a_{c}}$ is negative when $\rho_{c}$ is less than $0.661$. In Figs. \ref{dustScaleVsRho}.b and \ref{radScaleVsRho}.b, we show that $\ddot{a_{c}}$ with respect to $\rho_c$ for matter and radiation cases, respectively. As can be seen from the figures, $\ddot{a_{c}}$ is negative when $\ddot{a_{c}}<0.661$. In Figs. \ref{dustScaleVsRho}.a and \ref{radScaleVsRho}.a, $\ddot{a_{c}}=0$ is represented by an empty circle. For the matter-dominated universe, the right branch and the small portion of the left branch exhibit bouncing behavior. Interestingly, $\rho_{c}$ where $a_{c}$ is minimum and $\rho_{c}$ where $\ddot{a_{c}}=0$ are the same for the radiation dominated universe. Thus, the left branch in Fig. \ref{radScaleVsRho}.a corresponds to negative $\ddot{a_{c}}$ and does not exhibit the bouncing behaviour. The bouncing behaviour is only available in the right branch for the radiation case.
\begin{table}[]
\footnotesize
    \centering
\begin{tabular}{cccccc}
\hline 
 & $\alpha=0.2$ & $\alpha=0.4$ & $\alpha=0.6$ & $\alpha=0.8$ \tabularnewline
\hline 
\hline 
$a_{c}>0(k=1)$ & $\rho_{c}\in(0,6.462)$ & $\rho_{c}\in(0,3.565)$ & $\rho_{c}\in(0,2.573)$ & $\rho_{c}\in(0,2.065)$ \tabularnewline
\hline 
$\ddot{a_{c}}>0(k=1)$ & $\rho_{c}\in(2.22,6.462)$ & $\rho_{c}\in(1.262,3.565)$& $\rho_{c}\in(0.934,2.573)$ & $\rho_{c}\in(0.765,2.065)$ 
\tabularnewline
\hline 
$a_{c}, \ddot{a_{c}}>0 (k=-1)$ & $\rho_{c}>6.462$ & $\rho_{c}>3.565$ & $\rho_{c}>2.573$ & $\rho_{c}>2.065$ \tabularnewline
\hline
\end{tabular}
    \caption{The conditions $a_{c}>0$ and $\ddot{a}_c>0$ for the different values of $\alpha$. We take $\omega=0$.}
    \label{table1}
\end{table}
\begin{table}[]
\footnotesize
    \centering
\begin{tabular}{cccccc}
\hline 
 & $\alpha=0.2$ & $\alpha=0.4$ & $\alpha=0.6$ & $\alpha=0.8$ \tabularnewline
\hline 
\hline 
$a_{c}>0(k=1)$ & $\rho_{c}\in(0,4.391)$ & $\rho_{c}\in(0,2.462)$ & $\rho_{c}\in(0,1.8)$ & $\rho_{c}\in(0,1.46)$ \tabularnewline
\hline 
$\ddot{a_{c}}>0(k=1)$ & $\rho_{c}\in(2.22,4.391)$ & $\rho_{c}\in(1.262,2.462)$& $\rho_{c}\in(0.934,1.8)$ & $\rho_{c}\in(0.765,1.46)$ 
\tabularnewline
\hline
$a_{c}, \ddot{a_{c}}>0 (k=-1)$ & $\rho_{c}>4.391$ & $\rho_{c}>2.462$ & $\rho_{c}>1.8$ & $\rho_{c}>1.46$ \tabularnewline
\hline
\end{tabular}
    \caption{The conditions $a_{c}>0$ and $\ddot{a}_c>0$ for the different values of $\alpha$. We take  $\omega=1/3$.}
    \label{table2}
\end{table}

The case $k=-1$ clearly shows different behavior. In Figs. \ref{dustScaleVsRho} and \ref{radScaleVsRho}, red dashed line shows the case $k=-1$. As can be seen from Figs. \ref{dustScaleVsRho}.a and \ref{radScaleVsRho}.a, $a_{c}$ always decreases while $\rho_{c}$ increases. From Figs. \ref{dustScaleVsRho}.b and \ref{radScaleVsRho}.b, we see that $\ddot{a_{c}}$ always increases when $\rho_{c}$ increases. In contrast to the closed case ($k=1$), $\ddot{a_{c}}$ is always positive. Therefore, the whole branch of $a_{c}$ shows the bouncing behaviour for matter and radiation cases.

We only present the case $\alpha=1$ in the figures since similar behaviours can be seen for the different values of $\alpha$. However, the range of $\rho_{c}$ for the conditions $a_{c}>0$ and $\ddot{a_{c}}>0$ is presented in Tables \ref{table1} and \ref{table2} for the different values of $\alpha$. In Tables \ref{table3} and \ref{table4}, we also give the minimum $a_{c}$ for the different values of $\alpha$.  From Tables \ref{table3} and \ref{table4}, one can see that the minimum $a_{c}$ increases and $\rho_{c}$ decreases when $\alpha$ increases.

Finally, we consider the case $k=0$. We see that $a_{c}$ vanishes and the condition $a_{c}>0$ is not satisfied. Therefore, we conclude that the bouncing behaviour is absent for the case $k=0$. 
\begin{table}[]
    \centering
\begin{tabular}{cccccc}
\hline 
 & $\alpha=0.2$ & $\alpha=0.4$ & $\alpha=0.6$ & $\alpha=0.8$ \tabularnewline
\hline 
\hline 
$\rho_{c}$ & $3.263$ & $1.819$ & $1.326$ & $1.073$ \tabularnewline
\hline 
$a_{c}$ & $0.258$ & $0.337$& $0.388$ & $0.426$ 
\tabularnewline
\hline
\end{tabular}
    \caption{The minimum $a_{c}$ for the different values of $\alpha$. We take  $\omega=0$ and $k=1$.}
    \label{table3}
\end{table}
\begin{table}[]
    \centering
\begin{tabular}{cccccc}
\hline 
 & $\alpha=0.2$ & $\alpha=0.4$ & $\alpha=0.6$ & $\alpha=0.8$ \tabularnewline
\hline 
\hline 
$\rho_{c}$ & $2.22$ & $1.262$ & $0.934$ & $0.765$ \tabularnewline
\hline 
$a_{c}$ & $0.309$ & $0.399$& $0.455$ & $0.496$ 
\tabularnewline
\hline
\end{tabular}
    \caption{The minimum $a_{c}$ for the different values of $\alpha$. We take  $\omega=1/3$ and $k=1$.}
    \label{table4}
\end{table}

\section{Conclusions}
\label{sec4}

In this paper, employing the exponential corrected entropy \cite{Chatterjee2020}, we obtained the modified Friedmann equations from the first law of thermodynamics and Verlinde's entropic gravity. First, we obtained the exponential corrected Friedmann equation from the first law of thermodynamics at the apparent horizon \cite{Akbar2007}. We computed the deceleration parameter for the flat case $k=0$ and found that the expansion for radiation and matter dominated eras is more decelerated since exponential correction gives the positive contribution to $q$ parameter. Moreover, we investigated the validity of GSL and showed that the GSL is always satisfied for the exponential corrected Friedmann equations obtained from the first law.

Then, we derived the exponential corrected Friedmann equations from Verlinde's entropic gravity scenario \cite{Verlinde2011,Cai2010}. Similarly, we computed the deceleration parameter for the exponential corrected Friedmann equations obtained from  entropic gravity. Unlike the deceleration parameter (\ref{qMattAndRad}) from the first law, we found that the deceleration parameter (\ref{decelEntropicMatAndRad}) from  entropic gravity is positive for radiation and matter dominated eras when the conditions (\ref{rCondition}) and (\ref{mCondition}) are satisfied.

Finally, we numerically investigated the bouncing behaviour for the matter and radiation cases. The bouncing behaviour circumvents the initial singularity since the quantum gravity effects play an important role in the early stages of universe \cite{Pan2016,Jusufi2023}. We showed that the bouncing behaviour is available for the cases $k=1$ and $k=-1$. In Figs. \ref{dustScaleVsRho}.a and \ref{radScaleVsRho}.a, we showed that $a_{c}$ has two branches for the closed universe ($k=1$). For the matter dominated and closed universe, we concluded that the right branch and small portion of left branch in Fig. \ref{dustScaleVsRho}.a exhibit the bouncing behavior. For the radiation dominated and closed universe, we concluded that the right branch in Fig. \ref{radScaleVsRho}.a  only has the bouncing behaviour. For the open case ($k=-1$), we showed that $a_{c}$ has only one branch in Figs. \ref{dustScaleVsRho}.a and \ref{radScaleVsRho}.a. We concluded that whole branch of $a_{c}$ has bouncing behaviour since $\ddot{a_{c}}$ is always positive. For the flat case ($k=0$), $a_{c}$ vanishes. Thus, the flat universe does not show a bouncing behaviour.

\subsubsection*{Acknowledgments}
The authors are grateful to the anonymous reviewer for his/her helpful and constructive comments. This work was supported by Istanbul University Post-Doctoral Research Project: MAB-2021-38032.

\subsubsection*{Author Contributions}
Ö.Ö. wrote the paper, performed the calculations and prepared the figures. E. A. is the supervisor and checked the calculations. All authors reviewed the manuscript.

\subsubsection*{Funding}
Open access funding provided by the Scientific and Technological Research Council of Türkiye (TÜBiTAK).

\subsubsection*{Data Availability}
No data was used for the research described in the article.

\subsection*{Declarations}
\subsubsection*{Conflict of interest }
The authors declare that they have no conflict of interest.

\end{document}